\documentclass[12pt]{article}
\usepackage{amsmath}
\usepackage[dvips]{graphicx}
\usepackage{epsfig}
\usepackage{amsmath}
\usepackage{amssymb}
\usepackage{graphics,epstopdf}
\usepackage{amsfonts}
\usepackage{amsthm}
\usepackage[usenames]{color}

\definecolor{AV}{rgb}{0.65,0.0,0}
\definecolor{GC}{rgb}{0,0.0,0.65}
\definecolor{WS}{rgb}{0,0.65,0}

\usepackage[
      colorlinks=true,
      linkcolor=blue,
      urlcolor=blue,
      filecolor=blue,
      citecolor=blue,
      pdfstartview=FitV,
      pdftitle={},
      pdfauthor={},
      pdfsubject={},
      pdfkeywords={},
      pdfpagemode=None,
      bookmarksopen=true
]{hyperref}

\newcommand{\bm}{\begin{multiline}}
\newcommand{\beq}{\begin{equation}}
\newcommand{\eeq}{\end{equation}}
\newcommand{\beqs}{\begin{eqnarray}}
\newcommand{\eeqs}{\end{eqnarray}}

\begin{document}

\thispagestyle{empty}

\hfill{}

\hfill{}

\hfill{}

\vspace{32pt}

\begin{center}

\textbf{\Large  Charged black holes on Kaluza-Klein bubbles }

\vspace{48pt}

\textbf{Jutta Kunz,}\footnote{Email: \texttt{jutta.kunz@uni-oldenburg.de}}
\textbf{Petya G. Nedkova,}\footnote{Email: \texttt{pnedkova@phys.uni-sofia.bg}}
\textbf{Cristian Stelea,}\footnote{E-mail: \texttt{cristian.stelea@outlook.com}}

\vspace*{0.2cm}

\textit{$^{1,2}$Institut f\"ur Physik, Universit\"at Oldenburg,}\\[0pt]
\textit{D--26111 Oldenburg, Germany}\\[.5em]

\textit{$^3$Faculty of Physics, ``Alexandru Ioan Cuza" University}\\[0pt]
\textit{11 Blvd. Carol I, Iasi, 700506, Romania}\\[.5em]

\end{center}

\vspace{30pt}

\begin{abstract}

 We construct exact solutions of the Einstein-Maxwell-Dilaton field equations in five dimensions, which describe general configurations of charged and static black holes sitting on a Kaluza-Klein bubble. More specifically we discuss the configurations describing two black holes sitting on a Kaluza-Klein bubble and also the general charged static black Saturn balanced by a Kaluza-Klein bubble. A straightforward extension of the solution generating technique leads to a new solution describing the charged static black Saturn on the Taub-bolt instanton. We compute the conserved charges and investigate some of the thermodynamic properties of these systems.
\end{abstract}

\vspace{32pt}

\setcounter{footnote}{0}

\newpage

\section{Introduction}

In general, bubbles of nothing are smooth time dependent vacuum solutions of Einstein's field equations. Their defining feature is that they have a minimal area surface, with no space inside. The first example was provided by Witten as the end state of the decay of the Kaluza-Klein vacuum. More precisely, in the original Kaluza-Klein (KK) theory in five dimensions the vacuum state is not the five-dimensional Minkowski space, instead it is $M^4\times S^1$, the product of the four-dimensional Minkowski space with a KK circle $S^1$.
 Witten showed that the KK vacuum can be non-perturbatively unstable against decay to a time-dependent KK bubble \cite{Witten:1981gj}. By performing an analytical continuation of the five-dimensional Schwarzschild black hole, Witten obtained a Euclidean instanton which describes at time zero the nucleation of a minimal area $S^2$ ``bubble of nothing" where the KK circle smoothly pinches off. From a four-dimensional perspective this $S^2$ bubble has no interior, hence the name ``bubble of nothing". Once it forms, the bubble rapidly expands to null infinity, ``eating up" the space-time in the process. Fortunately, the production of KK bubbles is only allowed if one imposes anti-periodic boundary conditions for fermions around the KK circle. As such, the decay of the KK vacuum is forbidden in a theory with elementary fermions and when one imposes supersymmetric boundary conditions. More general time-dependent KK bubbles obtained by double analytical continuations of the Kerr solution have been studied in \cite{Aharony:2002cx}. A similar decay process in which a certain orbifold in Anti-de Sitter space-time decays to a bubble of nothing was found to exist by Balasubramanian \textit{et al.} in \cite{Balasubramanian:2005bg}. Bubbles of nothing obtained by double analytical continuations from general NUT-charged spaces have been studied in \cite{Ghezelbash:2002xt,Astefanesei:2005yj}. By using a cleverly chosen ansatz, Copsey was able to show the existence of bubbles of nothing in asymptotically flat spaces in \cite{Copsey:2006qb} and later generalized them in Anti-de Sitter context in \cite{Copsey:2007hw}. Locally these bubbles look like KK bubbles, having a minimal area sphere $S^2$ where a circle pinches off, however, they are asymptotically globally flat/Anti-de Sitter.

More recently, there has been renewed interest in bubbles of nothing since it was pointed out by Horowitz that they can provide new endpoints for the Hawking evaporation of charged black strings \cite{Horowitz:2005vp}. For example, in the presence of a charged black string the KK compactification radius of the $S^1$ circle varies with the radial coordinate. For certain values of the string charges, this radius can be made to vary arbitrarily slowly and it can reach the string scale near and outside the black string horizon. In this region, a tachyon will appear which will cause the KK circle to pinch off and form a KK bubble of nothing. Unlike Witten's bubble, the bubbles of nothing formed in the presence of the black string can carry charges and moreover, in certain conditions they cannot be static, instead they have to expand to infinity. Bubbles carrying magnetic charges have been discussed in \cite{Stotyn:2011tv} where it was found that the addition of a topological magnetic charge can locally stabilize the KK vacuum.

While the KK bubbles of nothing that describe the non-perturbative decay of the KK vacuum are highly dynamical, there are also known solutions which describe static KK bubbles. For example, a static KK bubble in five dimensions can be easily obtained by taking the product of the four-dimensional Euclidean Schwarzschild solution with a trivial time direction. Since the Euclidian Schwarzschild solution is an example of an asymptotically flat gravitational instanton in four dimensions, then the five-dimensional solutions that describe black holes on static KK bubble backgrounds fall into the class of solutions recently introduced by Chen and Teo in \cite{Chen:2010ih}. More generally, they correspond to the so-called black holes on gravitational instantons since in absence of black holes the background geometry is a direct product of a trivial time direction with a four-dimensional Ricci flat instanton with $U(1)\times U(1)$ symmetry \cite{Chen:2010zu}. The construction and physical interpretation of such solutions has been greatly facilitated by the use of the generalized Weyl formalism introduced in \cite{Emparan:2001wk,Harmark:2004rm}. For example, in four dimensions a solution describing two colliding KK bubbles has been studied in \cite{Horowitz:2002cx}. The first solution describing a single black hole on a KK bubble was analyzed in \cite{Emparan:2001wk} and subsequently generalized to the case of two black holes on a KK bubble in \cite{Elvang:2002br}. Sequences of static black holes on KK bubbles have been considered in \cite{Elvang:2004iz,Harmark:2007md}.  One of the most surprising results of these studies was that, for instance, two large black holes can be held in static equilibrium by a single KK bubble, even if the separation distance between the black holes is very small compared to their size. This means that one can construct multi-black holes on KK bubble spacetimes that do not have conical singularities, that is, the black holes are kept in static equilibrium. One can of course add rotation to these systems and describe rotating black holes on KK bubbles. For example, a rotating black ring on a KK bubble has been studied in \cite{Nedkova:2010gn}. This solution is similar to the rotating asymptotically flat black ring found by Emparan and Reall \cite{Emparan:2001wn}. Rotating generalizations of the solution describing two black holes on a KK bubble have been constructed in \cite{Iguchi:2007xs,Tomizawa:2007mz}. Charged generalizations of these vacuum solutions to describe charged black holes on KK bubbles have been studied in \cite{Kunz:2008rs,Yazadjiev:2009gr,Yazadjiev:2009nm}.

The aim of this paper is to further extend and study the class of solutions describing charged black holes on KK bubbles. In the first part of this article we shall construct two general solutions which describe two charged black holes/rings on a KK bubble. The first solution is a generalization of the electrically charged double black hole constructed in \cite{Kunz:2008rs}. More precisely, for the solution derived in \cite{Kunz:2008rs} the charges and masses of the two black holes cannot take independent values, since, as an artefact of the particular solution-generating technique used to generate them, the mass to charge ratios take the same value for each black hole in that system. By contrast, in our solution-generating technique the final solution is described by five independent parameters, corresponding to the values of the masses and charges of the black holes and the separating distance between them (equivalently, the rod length of the KK bubble in between the black holes). The second solution that we discuss is a KK analog of the black Saturn constructed by Elvang and Figueras in \cite{Elvang:2007rd}. The ring of the Saturn is now placed on two KK bubbles while the black hole is placed at the other end of one of the bubbles. The vacuum version of the black Saturn on the KK bubbles is included in the class of solutions studied in \cite{Elvang:2004iz} and, in our case, it is simply obtained by setting the electric charges equal to zero. In the generalized Weyl formalism, when discussing the rod structure of a given solution, one can interpret a finite rod along a spacelike direction as representing a KK bubble. This is why black holes on the Schwarzschild instanton correspond to black holes on KK bubbles. Such finite spacelike rods also appear in the rod structures of the Euclidean Kerr and the Taub-bolt instanton \cite{Chen:2010zu} and, in this sense, one could interpret black holes on these instantons as corresponding to black holes on generalized KK bubbles, with different asymptotic structures. For example, a black hole on the Euclidean Kerr instanton has been derived in \cite{Chen:2010ih}. Black holes on the Taub-bolt instanton have been studied in \cite{Chen:2010ih,Stelea:2012ph,Stelea:2012xg,Nedkova:2011hx}. Based on these considerations, in the second part of this paper we derive a more general solution that describes a black Saturn on the Taub-bolt instanton in five dimensions. Most of the ingredients used to construct this solution have been already computed in the derivation of the black Saturn on the KK bubbles and it will be easy to see that for certain values of the parameters the black Saturn on the KK bubbles can be obtained as a particular case.

The structure of our paper is as follows: in the next section we describe the solution-generating technique and introduce the seed solution from four dimensions. In section $3$ we construct two new solutions: the charged double-black holes on a KK bubble and also the general charged black Saturn on KK bubbles. In section $4$ we derive the charged black Saturn solution on the Taub-bolt instanton. Finally, in section $5$  we compute the conserved quantities and investigate some thermodynamic properties of these solutions. The final section is dedicated to conclusions.

\section{The solution generating technique}

The main idea of the solution-generating technique that we shall put to use is to map a general static electrically charged axisymmetric solution of Einstein-Maxwell theory in four dimensions to a five-dimensional static electrically charged axisymmetric solution of the Einstein-Maxwell-Dilaton (EMD) theory with arbitrary coupling of the dilaton to the electromagnetic field. To this end one performs a dimensional reduction of both theories down to three dimensions and, after a mapping of the corresponding scalar fields and electromagnetic potentials of each theory, one is able to bypass the actual solving of the field equations by algebraically mapping solutions of one theory to the other. This solution generating technique has been previously used in \cite{Chng:2008sr,Stelea:2011jm} to construct multi-black hole systems in five-dimensional asymptotically flat backgrounds. However, for simplicity, in what follows we shall focus only on the Einstein-Maxwell theory and, therefore, we set to zero the coupling of the dilaton field to the Maxwell field and further set the dilaton field to zero.

To obtain general charged and static configurations of double black holes on KK bubbles, we shall make use of the four-dimensional double Reissner-Nordstr\"om solution in the parametrization given recently by Manko in \cite{Manko:2007hi} with metric

\begin{eqnarray}
&&ds^2 = -\tilde f dt^2 + \tilde f^{-1}\left[e^{2\tilde\mu}(d\rho^2 + dz^2) + \rho^2 d\varphi\right], \nonumber \\[2mm]
&&\tilde{f}=\frac{\tilde{A}^{2}-\tilde{B}^{2}+\tilde{C}^{2}}{(\tilde{A}+\tilde{B})^{2}},~~~~~~e^{2\tilde{\mu}}=\frac{%
\tilde{A}^{2}-\tilde{B}^{2}+\tilde{C}^{2}}{16\sigma _{1}^{2}\sigma _{2}^{2}(\nu
+2k)^{2}r_{1}r_{2}r_{3}r_{4}},  \label{Manko}
\end{eqnarray}%
and electromagnetic field
\begin{eqnarray}
\tilde{F} = d\Psi\wedge dt, \quad~~~~~~~\Psi=-\frac{2\tilde{C}}{\tilde{A}+\tilde{B}}
\end{eqnarray}
We have used the following notations
\begin{eqnarray}
\tilde{A} &=&\sigma _{1}\sigma _{2}[\nu
(r_{1}+r_{2})(r_{3}+r_{4})+4k(r_{1}r_{2}+r_{3}r_{4})]-(\mu ^{2}\nu
-2k^{2})(r_{1}-r_{2})(r_{3}-r_{4}),  \notag \\
\tilde{B} &=&2\sigma _{1}\sigma _{2}[(\nu M_{1}+2kM_{2})(r_{1}+r_{2})+(\nu
M_{2}+2kM_{1})(r_{3}+r_{4})]  \notag \\
&&-2\sigma _{1}[\nu \mu (Q_{2}+\mu )+2k(RM_{2}+\mu Q_{1}-\mu
^{2})](r_{1}-r_{2})  \notag \\
&&-2\sigma _{2}[\nu \mu (Q_{1}-\mu )-2k(RM_{1}-\mu Q_{2}-\mu
^{2})](r_{3}-r_{4}),  \notag \\
\tilde{C} &=&2\sigma _{1}\sigma _{2}\{[\nu (Q_{1}-\mu )+2k(Q_{2}+\mu
)](r_{1}+r_{2})+[\nu (Q_{2}+\mu )+2k(Q_{1}-\mu )](r_{3}+r_{4})\}  \notag \\
&&-2\sigma _{1}[\mu \nu M_{2}+2k(\mu M_{1}+RQ_{2}+\mu R)](r_{1}-r_{2})
\notag \\
&&-2\sigma _{2}[\mu \nu M_{1}+2k(\mu M_{2}-RQ_{1}+\mu R)](r_{3}-r_{4}),
\end{eqnarray}%
where the solution parameters are given by
\begin{eqnarray}
\nu &=&R^{2}-\sigma _{1}^{2}-\sigma _{2}^{2}+2\mu
^{2},~~~~~~~k=M_{1}M_{2}-(Q_{1}-\mu )(Q_{2}+\mu ),  \notag \\
\sigma _{1}^{2} &=&M_{1}^{2}-Q_{1}^{2}+2\mu Q_{1},~~~~~~~\sigma
_{2}^{2}=M_{2}^{2}-Q_{2}^{2}-2\mu Q_{2},~~~~~~~\mu =\frac{%
M_{2}Q_{1}-M_{1}Q_{2}}{M_{1}+M_{2}+R},
\end{eqnarray}%
and $r_{i}=\sqrt{\rho ^{2}+(z - a_i)^{2}}$, $i=1..4$. The parameters $a_i$ correspond  to the turning points in the rod structure of the seed solution, and they take the values
\begin{equation}
a_1 = \frac{R}{2}+\sigma _{2},~~~~~a_2 = \frac{R}{2}-\sigma
_{2},~~~~~a_3 = -\frac{R}{2}+\sigma _{1},~~~~~a_4 = -\frac{R}{2}
-\sigma _{1},
\label{ai}
\end{equation}%
It is assumed that $R>0$, $\sigma_i>0$, and the following order is satisfied $a_4<a_3<a_2<a_1$. The solution is parameterized by five independent parameters, and they can be chosen as the intrinsic masses of the two Reissner-Nordstr\"{o}m black holes $M_{1,2}$, their local charges $Q_{1,2}$ and the distance $R$ separating them. For a detailed discussion of its properties we refer the reader to \cite{Manko:2007hi,Manko:2008gb} and the references therein. Note that, in general, the function $e^{2\tilde{\mu}}$ can be determined up to a
constant and its precise numerical value has been fixed here by allowing the
presence of conical singularities only in the portion in between the black
holes along the $\varphi $ axis. Consequently one has:
\begin{equation}
e^{2\tilde{\mu}}|_{\rho =0}=\left( \frac{\nu -2k}{\nu +2k}\right) ^{2},
\label{strutManko}
\end{equation}%
for $-R/2+\sigma _{1}<z<R/2-\sigma _{2}$ and $e^{2\tilde{\mu}}|_{\rho =0}=1$
elsewhere.

In what follows we assume that all the functions involved in our solutions depend only on the canonical coordinates $\rho$ and $z$. The final solution of the Einstein-Maxwell system in five dimensions with the Lagrangian
\begin{eqnarray}
\mathcal{L}_{5}=\sqrt{-g}\left[R-\frac{1}{4}F_{(2)}^2\right]
\label{EMDaction5d}
\end{eqnarray}
where $F_{(2)}=dA_{(1)t}\wedge dt$, can be written as:
\beqs
ds^2&=&-\tilde{f}dt^2+Fd\chi^2+\frac{1}{G}\big[e^{2\Gamma}(d\rho^2+dz^2)+\rho^2 d\varphi^2\big],~~~~~~~A_{(1)t}=\frac{\sqrt{3}}{2}\Psi,
\label{finalKK}
\eeqs
where we defined the following functions:
\beqs
F&=&\tilde{f}^{-\frac{1}{2}}e^{2h},~~~~G=\tilde{f}^{\frac{1}{2}}e^{2h},~~~~e^{2\Gamma}=e^{\frac{3\tilde{\mu}}{2}+2\gamma}.
\label{functions}
\eeqs
The coordinate $t$ parameterizes the asymptotically timelike Killing field, and the periodic coordinates $\varphi$ and $\chi$ correspond to the spacelike Killing fields. Since in this paper we are interested in solutions with Kaluza-Klein asymptotic, we assume hereafter that the coordinate $\chi$ parameterizes the KK circle at infinity.
The function $h$ appearing in the solution is an arbitrary harmonic function\footnote{That is, it satisfies the equation $\nabla^2h=\frac{\partial^2h}{\partial\rho^2}+\frac{1}{\rho}\frac{\partial h}{\partial\rho}+\frac{\partial^2h}{\partial z^2}=0.$}, which can be chosen at will. Note that the presence of $h$ allows us to modify the rod structure of the final solution along the $\chi$ and $\varphi$ directions. Indeed, by carefully choosing the form of $h$, one can construct the appropriate rod structures to describe the desired configurations involving black holes sitting on KK bubbles. Finally, once the form of $h$ has been specified for a particular solution, the remaining function  $\gamma$ can be obtained by simple quadratures by using the equations:
\begin{eqnarray}  \label{gammap1a}
\partial_\rho{\gamma}&=&\rho[(\partial_\rho h)^2-(\partial_z h)^2],~~~~~~~
\partial_z{\gamma}=2\rho(\partial_\rho h)(\partial_z h).
\end{eqnarray}

\section{Black holes on a Kaluza-Klein bubble}

In this section we construct two new solutions describing charged black holes on KK bubbles. The first solution is the most general static solution involving two charged black holes on a KK bubble, while the second solution will correspond to the charged static black Saturn on KK bubbles.

\subsection{Charged double black holes on a KK bubble}

%%%%%%%%%%%%%%%%%%%%%%%%%%%%%%%%%%%%%%%%%%
\begin{figure}[tbp]
\par
\begin{center}
\includegraphics{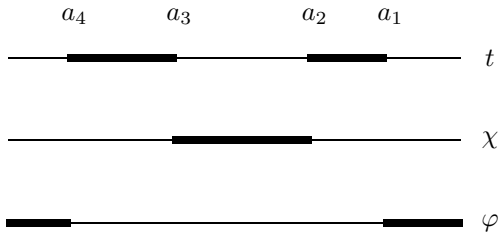}
\end{center}
\caption{Rod structure of the double black hole on a KK bubble.}
\label{2BHsKK}
\end{figure}
%%%%%%%%%%%%%%%%%%%%%%%%%%%%%%%%%%%%%%%%%%

The rod structure of a system of two black holes on a KK bubble is depicted in Figure \ref{2BHsKK}. We should construct the harmonic function $h$ in such a way that the metric ($\ref{finalKK}$) describes a solution possessing two horizons with spherical topology separated by a KK bubble. One possible form is
\beqs
e^{2h}&=&\frac{\sqrt{(r_1+\zeta_1)(r_2+\zeta_2)}}{\sqrt{(r_3+\zeta_3)(r_4+\zeta_4)}},
\eeqs
where $\zeta_i = z-a_i$. Then, if we integrate eq. (\ref{gammap1a}) we obtain the remaining metric function
\beqs
e^{2\gamma}&=&\frac{2}{K_0}\left(\frac{Y_{13}Y_{14}Y_{23}Y_{24}}{r_1r_2r_3r_4Y_{12}Y_{34}}\right)^{\frac{1}{4}},
\eeqs
where we use the notation $Y_{ij}=r_ir_j+\zeta_i\zeta_j+\rho^2$, and $K_0$ is an integration constant. We can show that the constructed solution (\ref{finalKK}) with these explicit metric functions is characterized by the rod structure depicted in Fig. \ref{2BHsKK}.

Following the procedure given in \cite{Harmark:2004rm,Chen:2010zu}, one deduces that the rod structure contains four turning points that divide the $z$-axis into five rods. They can be described in the following way, specifying the rod direction vectors  with respect to a basis of Killing fields $\{\frac{\partial}{\partial t}, \frac{\partial}{\partial\chi}, \frac{\partial}{\partial \phi}\}$, and  normalizing them to the surface gravity of each fixed point set:
\begin{itemize}
\item \textbf{Rod 1} - For $z<a_4$ one has a semi-infinite spacelike rod with normalized direction
\beqs
l_1=\sqrt{\frac{2\sqrt{2}}{K_0}}(0,0,1).
\eeqs
\item \textbf{Rod 2} - For $a_4<z<a_3$ one has a finite timelike rod that corresponds to a black hole horizon  with $S^3$ topology. Its normalized rod direction is given by
$l_2=\frac{1}{k_{H_1}}(1,0,0)$, where
\beqs
k_{H_1}=\sqrt{\frac{K_0}{2\sqrt{2}}}p_1\bigg[\left(\rho\tilde{f}^{-1}e^{\tilde{\mu}}\right)|_{H_1}\bigg]^{-\frac{3}{4}}
\label{haw1}
\eeqs
 is the surface gravity on the black hole horizon represented by this rod.
\item \textbf{Rod 3} - For $a_3<z<a_2$ one has a finite spacelike rod with normalized direction
$l_3=\frac{1}{k_{B}}(0,1,0)$,
where
\beqs
k_{B}=\frac{1}{2}\sqrt{\frac{K_0}{2\sqrt{2}}}\left|\frac{\nu+2k}{\nu-2k}\right|^{\frac{3}{4}}
\big[((R+\sigma_2)^2-\sigma_1^2)((R-\sigma_2)^2-\sigma_1^2)\big]^{-\frac{1}{4}}.
\eeqs
It corresponds to a KK bubble with surface gravity $k_{B}$.
\item \textbf{Rod 4} - For $a_2<z<a_1$ one has a finite timelike rod, corresponding to a second black hole horizon with $S^3$ topology. Its normalized rod direction is found to be
$l_4=\frac{1}{k_{H_2}}(1,0,0)$, where
\beqs
k_{H_2}=\sqrt{\frac{K_0}{2\sqrt{2}}}p_2\bigg[\left(\rho\tilde{f}^{-1}e^{\tilde{\mu}}\right)|_{H_2}\bigg]^{-\frac{3}{4}}
\label{haw2}
\eeqs
is the surface gravity of the black hole horizon corresponding to this rod.
\item \textbf{Rod 5} - for $z>a_1$ one has a semi-infinite spacelike rod with normalized direction
\beqs
l_5=\sqrt{\frac{2\sqrt{2}}{K_0}}(0,0,1).
\eeqs
\end{itemize}

Here we defined the following quantities:
\beqs
p_1&=&\left(\frac{\sigma_1}{(R+\sigma_1)^2-\sigma_2^2}\right)^{\frac{1}{4}},~~~~~p_2=\left(\frac{\sigma_2}{(R+\sigma_2)^2-\sigma_1^2}\right)^{\frac{1}{4}},
\label{pi}
\eeqs
while for each black hole horizon one has \cite{Manko:2008gb}:
\beqs
\left(\rho\tilde{f}^{-1}e^{\tilde{\mu}}\right)|_{H_1}&=&\frac{\big[(R+M_1+M_2)(M_1+\sigma_1)-Q_1(Q_1+Q_2)\big]^2}{\sigma_1[(R+\sigma_1)^2-\sigma_2^2]},\nonumber\\
\left(\rho\tilde{f}^{-1}e^{\tilde{\mu}}\right)|_{H_2}&=&\frac{\big[(R+M_1+M_2)(M_2+\sigma_2)-Q_2(Q_1+Q_2)\big]^2}{\sigma_2[(R+\sigma_2)^2-\sigma_1^2]}.
\label{rhofmu}
\eeqs
The integration constant $K_0$ should be equal to $2\sqrt{2}$ in order for the solution to be asymptotically  Kaluza-Klein. Then it is clear that the rod structure of this solution is the one depicted in (\ref{2BHsKK}) as advertised. Note that $k_B$ is the so-called surface gravity of the KK bubble as defined in \cite{Kastor:2008wd} and, for a smooth KK bubble, its relation to the length at infinity of the KK circle is given by:
\beqs
k_B&=&\frac{2\pi}{L}.
\eeqs
This fixes the value of $L$ to be:
\beqs
L&=&4\pi\left([(R+\sigma_2)^2-\sigma_1^2][(R-\sigma_2)^2-\sigma_1^2]\right)^{\frac{1}{4}}\left|\frac{\nu-2\mu}{\nu+2\mu}\right|^{\frac{3}{4}}.
\eeqs

\subsection{The charged black Saturn on the KK bubbles}

 %%%%%%%%%%%%%%%%%%%%%%%%%%%%%%%%%%%%%%%%%%
\begin{figure}[tbp]
\par
\begin{center}
\includegraphics{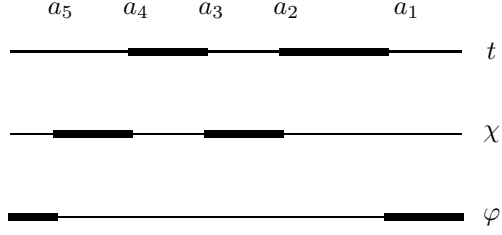}
\end{center}
\caption{Rod structure of the black Saturn on KK bubbles.}
\label{SaturnKK}
\end{figure}
%%%%%%%%%%%%%%%%%%%%%%%%%%%%%%%%%%%%%%%%%%

To construct a solution which describes a static charged black Saturn on a KK bubble one should choose another form of the harmonic function $h$, which can be guessed by the rod structure of the solution.  Since one of the black hole horizons should possess a ring topology, the rod corresponding to it should be surrounded with two finite rods along the Killing field $\partial/\partial\chi$. The other  black hole horizon should be topologically $S^3$, so it should intersect with a rod directed along $\partial/\partial\chi$, and a rod directed along the other spacelike Killing field $\partial/\partial\varphi$. The solution should be asymptotically Kaluza-Klein, which further means that the two semi-infinite rods should be directed along the Killing field $\partial/\partial\varphi$, which is not associated with the compact dimension. All these considerations imply that the rod structure of the static black Saturn on KK bubbles can be described by Figure \ref{SaturnKK}. To obtain this rod structure  the harmonic function $h$ should possess the form
\beqs
e^{2h}&=&\sqrt{\frac{(r_1+\zeta_1)(r_2+\zeta_2)(r_4+\zeta_4)}{r_3+\zeta_3}}\frac{1}{r_5+\zeta_5}.
\label{hSaturnKK}
\eeqs
The metric function ã is found afterwards by integrating (\ref{gammap1a})
\beqs
e^{2\gamma}&=&\frac{\sqrt{2}}{K_0}\left(\frac{Y_{15}Y_{25}Y_{45}}{r_5^2Y_{35}}\right)^{\frac{1}{2}}\left(\frac{Y_{13}Y_{23}Y_{34}}{r_1r_2r_3r_4Y_{12}Y_{14}Y_{24}}\right)^{\frac{1}{4}},
\label{gammaSaturnKK}
\eeqs
where $K_0$ is an arbitrary constant. We shall show that the rod structure of the constructed solution coincides with that depicted in Figure \ref{SaturnKK}. From the explicit form of the solution  we obtain that the parameters denoting the rod endpoints $a_i$,  $i =1..4$, coincide with the corresponding values for the seed solution given by ($\ref{ai}$). An additional turning point is introduced at $z=a_5$, which is originally a free parameter. It is convenient to be chosen as $a_5=-\frac{R}{2}-\sigma$, where $\sigma>\sigma_1$. The six rods constituting the rod structure of the new solution can be described in the following way, where the rod direction vectors are given with respect to a basis of Killing fields $\{\frac{\partial}{\partial t}, \frac{\partial}{\partial\chi}, \frac{\partial}{\partial \phi}\}$, and they are also normalized to the surface gravity of each fixed point set:
\begin{itemize}
\item \textbf{Rod 1} - For $z<a_5$ one has a semi-infinite spacelike rod with normalized direction
\beqs
l_1=\sqrt{\frac{2\sqrt{2}}{K_0}}(0,0,1).
\eeqs
\item \textbf{Rod 2} - For $a_5<z<a_4$ one has a finite spacelike rod with normalized direction
$l_2=\frac{1}{k_{B_1}}(0,1,0)$,  where
\beqs
k_{B_1}=\frac{1}{2}\left[\frac{\sigma + \sigma_1}{\left(\sigma-\sigma_1\right)\left((R + \sigma)^2 - \sigma_2^2\right)}\right]^{1\over2}.
\eeqs
It represents a KK bubble with surface gravity $k_{B_1}$.
\item \textbf{Rod 3} - For $a_4<z<a_3$ one has a finite timelike rod that corresponds to a black hole horizon with ring topology ($S^2\times S^1$). It possesses normalized  direction given by
    $l_3=\frac{1}{k_{H_1}}(1,0,0)$, where
\beqs
k_{H_1}=\sqrt{\frac{K_0}{2\sqrt{2}}}p_1\bigg[\left(\rho\tilde{f}^{-1}e^{\tilde{\mu}}\right)|_{H_1}\bigg]^{-\frac{3}{4}}
\eeqs
is the surface gravity of the corresponding horizon.
\item \textbf{Rod 4} - For $a_3<z<a_2$ one has a finite spacelike rod with  normalized direction
$l_4=\frac{1}{k_{B_2}}(0,1,0)$, where
\beqs
k_{B_2}= \frac{1}{2}\left|\frac{\nu + 2k}{\nu-2k}\right|^{3\over4}\left[ (R+\sigma)^2 - \sigma_2^2\right]^{-{1\over2}}\left[\frac{(R+\sigma_1)^2 -\sigma_2^2}{(R-\sigma_1)^2 -\sigma_2^2}\right]^{1\over4}.
\eeqs
It represents a second KK bubble with surface gravity $k_{B_2}$.
\item \textbf{Rod 5} - For $a_2<z<a_1$ one has a finite timelike rod, corresponding to a second black hole horizon with $S^3$ topology. It possesses normalized  direction
$l_5=\frac{1}{k_{H_2}}(1,0,0)$, where
\beqs
k_{H_2}=\sqrt{\frac{K_0}{2\sqrt{2}}}p_2\bigg[\left(\rho\tilde{f}^{-1}e^{\tilde{\mu}}\right)|_{H_2}\bigg]^{-\frac{3}{4}}
\eeqs
is the surface gravity of the corresponding horizon.
\item \textbf{Rod 6} - for $z>a_1$ one has a semi-infinite spacelike rod with normalized direction
\beqs
l_6=\sqrt{\frac{2\sqrt{2}}{K_0}}(0,0,1).
\eeqs
\end{itemize}

The expressions of the surface gravity for each black hole include a pair of constants $p_i$, $i=1,2$. They characterize the expansion of the metric function $e^{2h-2\gamma}$ near each black hole horizon
\beqs
e^{2h-2\gamma}|_{H_i}&=&(p_i)^2\sqrt{\rho}+{\cal O}(\rho),
\eeqs
and in our case they can be computed using the formulae  (\ref{hSaturnKK}) and (\ref{gammaSaturnKK}) as
\beqs\label{Pi_KK}
p_1 &=& \frac{1}{\sqrt{2}}\left[\frac{\sigma_1 + \sigma}{(R + \sigma)^2 - \sigma_2^2}\right]^{1\over2}\left[\frac{(R+\sigma_1)^2 - \sigma_2^2}{\sigma_1}\right]^{1\over4}, \nonumber \\[2mm]
p_2 &=& \left(R + \sigma + \sigma_2\right)^{-{1\over2}}\left[\frac{\sigma_2\left(R + \sigma_2 + \sigma_1\right)}{R + \sigma_2 - \sigma_1}\right]^{1\over4}.
\eeqs

The integration constant $K_0$ should have the value $K_0=2\sqrt{2}$ in order for the solution to be asymptotically  Kaluza-Klein. Other parameters are  further restricted by the requirement that conical singularities do not occur. The rod structure of the solution contains two rods directed along the Killing field $\partial/\partial\chi$, which correspond to the Kaluza-Klein bubbles. Consequently, to avoid conical singularities the periods of the orbits of $\partial/\partial\chi$ in the vicinity of the bubbles $\Delta\chi_{B_1}$ and $\Delta\chi_{B_2}$ should coincide with the length of the compact dimension at infinity $L$

\beqs
\Delta\chi_{B_1} = \frac{2\pi}{k_{B_1}} = L, \quad~~~ \Delta\chi_{B_2} = \frac{2\pi}{k_{B_2}} = L.
\eeqs

Considering the expressions for the surface gravities on the bubbles, we obtain that the two equalities are compatible if the following condition is satisfied

\beqs
\frac{\sigma-\sigma_1}{\sigma + \sigma_1} = \left|\frac{\nu - 2k}{\nu+2k}\right|^{3\over2}\left[\frac{(R-\sigma_1)^2 -\sigma_2^2}{(R+\sigma_1)^2 -\sigma_2^2}\right]^{1\over2}.
\eeqs
If the rod structure of the seed solution is considered fixed, this is basically a restriction on the length of the first bubble rod $a_5 < z < a_4$, or equivalently, on the value of the parameter $\sigma$. Finally we should note that in the limit $\sigma = \sigma_1$ the solution reduces to the double black holes on KK bubble, described in the previous section.

\section{The static charged black Saturn on the Taub-bolt instanton}

As we have mentioned in the introduction, the Taub-bolt instanton is another four-dimensional gravitational instanton that can be used as a background for black hole solutions in five dimensions. For example, a solution describing a single black hole on the Taub-bolt instanton has been derived in \cite{Chen:2010ih}. The black ring on the Taub-bolt instanton was constructed and studied in \cite{Stelea:2012xg}, while a system of two charged black holes was recently studied in \cite{Stelea:2012ph}. The Taub-bolt instanton contains a fixed point set of the spacelike Killing field associated with the compact dimension, which represents a regular surface with $S^2$ topology, also called a bolt \cite{Gibbons:1979c}. In this way it resembles the KK bubble. However, the asymptotic structure of the Taub-bolt instanton is different since it is only locally Kaluza-Klein ($M^4\times S^1$), but globally the spatial cross-sections at  infinity represent a more general fibre bundle. The solution is characterized by a NUT charge which is proportional to the first Chern class of the fibre bundle at infinity, and it contains the Kaluza-Klein bubble as a limit case when the NUT charge vanishes.

The solutions representing black holes on the Taub-bolt instanton obtained so far possess one important feature. In resemblance to the black hole and KK bubbles configurations they can be made completely regular for certain values of the parameters, so no conical singularities occur, and the black hole system is balanced. In this section we shall construct a more general exact solution, which describes a charged black Saturn on the Taub-bolt instanton. We shall show that for specific values of the parameters the static black Saturn on the Taub-bolt can be in equilibrium as well.

The details of the solution generating technique that we use have been previously presented in \cite{Stelea:2009ur} so that here we only give the final formulae. Since one wants to construct a solution describing two charged black holes/rings in five dimensions,  the starting point of the solution generating technique is again the charged four-dimensional double black hole solution given in (\ref{Manko}). Following the results in \cite{Stelea:2009ur}, one can write the final solution in the form:
\beqs
ds^2&=&-\tilde{f}dt^2+\frac{F}{\Sigma}(d\chi+\omega d\varphi)^2+\frac{\Sigma}{G}\big[e^{2\Gamma}(d\rho^2+dz^2)+\rho^2 d\varphi^2\big],~~~~~~~A_{(1)t}=\frac{\sqrt{3}}{2}\Psi,
\eeqs
where we defined the following functions:
\beqs
F&=&\tilde{f}^{-\frac{1}{2}}e^{2h},~~~~G=\tilde{f}^{\frac{1}{2}}e^{2h},~~~~\Sigma=A^2-C^2e^{4h},~~~~e^{2\Gamma}=e^{\frac{3\tilde{\mu}}{2}+2\gamma},~~~~\omega=4ACH.
\label{functionsTB}
\eeqs
Note that $A$ and $C$ are arbitrary constants, $h$ is again a harmonic function to be chosen at will, while the function $\gamma$ can be easily derived by integrating (\ref{gammap1a}).
In addition, the function $H$ is the so-called `dual' of $h$ and it is a solution of the following equation:
\beqs
dH&=&\rho(\partial_{\rho}h dz-\partial_zhd\rho).
\eeqs
Therefore, once the harmonic function $h$ is determined, to obtain the final solution one has to integrate (\ref{gammap1a}) and compute the dual of $h$. In order to have the right asymptotic at infinity, the values of the constants $A$ and $C$ should be chosen such that $A^2-C^2=1$.

To construct a black Saturn solution, which describes a black ring around a spherical black hole on the Taub-bolt instanton, it turns out that one has to pick the same harmonic function that was used to construct the black Saturn on KK bubbles. Namely, $h$ should be given by (\ref{hSaturnKK}), and $\gamma$ coincides with (\ref{gammaSaturnKK}). The dual of $h$ is easy to evaluate:\footnote{In general, the dual of $\frac{1}{2}\ln(r_i+\zeta_i)$ is $-\frac{1}{2}(r_i-\zeta_i)$, while the dual of $\frac{1}{2}\ln(r_i-\zeta_i)$ is $-\frac{1}{2}(r_i+\zeta_i)$, where $r_i=\sqrt{\rho^2+\zeta_i^2}$, $\zeta_i=z-a_i$ and $a_i$ is a constant.}
\beqs
H&=&\frac{1}{4}\big[r_3+2r_5-(r_1+r_2+r_4)\big],
\eeqs
up to a constant term.

To prove that the final solution describes indeed a static black Saturn on the Taub-bolt instanton we shall compute the rod structure of this solution. As in the case of the static black Saturn on KK bubbles, there are five turning points such that $a_i$, $i=1..4$, coincide with the rod endpoints of the seed solution ($\ref{ai}$),  and $a_5$ can be chosen as $a_5=-\frac{R}{2}-\sigma$, where $\sigma>\sigma_1$. They divide the $z$ axis into six rods which can be described in the following way, where the rod direction vectors are given with respect to a basis of Killing fields $\{\frac{\partial}{\partial t}, \frac{\partial}{\partial\chi}, \frac{\partial}{\partial \phi}\}$, and they are also normalized to the surface gravity of the corresponding fixed point set.
\begin{itemize}
\item \textbf{Rod 1} - For $z<a_5$ one has a semi-infinite spacelike rod with normalized direction
\beqs
l_1=\sqrt{\frac{2\sqrt{2}}{K_0}}(0,2AC(\sigma-\sigma_1+R),1).
\eeqs
\item \textbf{Rod 2} - For $a_5<z<a_4$ one has a finite spacelike rod with normalized direction
$l_2=\frac{1}{k_{B1}}(0,1,0)$, where
\beqs
k_{B_1}=\frac{1}{2A^2}\left[\frac{\sigma + \sigma_1}{\left(\sigma-\sigma_1\right)\left((R + \sigma)^2 - \sigma_2^2\right)}\right]^{1\over2}
\eeqs
It corresponds to a bolt, according to the classification in \cite{Gibbons:1979c}, with surface gravity $k_{B_1}$.
\item \textbf{Rod 3} - For $a_4<z<a_3$ one has a finite timelike rod that corresponds to a horizon with ring topology ($S^2\times S^1$). Its normalized rod direction is given by
$l_3=\frac{1}{k_{H_1}}(1,0,0)$, where
\beqs
k_{H_1}=\sqrt{\frac{K_0}{2\sqrt{2}}}\frac{p_1}{A}\bigg[\left(\rho\tilde{f}^{-1}e^{\tilde{\mu}}\right)|_{H_1}\bigg]^{-\frac{3}{4}}
\eeqs
is the surface gravity on the horizon.
\item \textbf{Rod 4} - For $a_3<z<a_2$ one has a finite spacelike rod with normalized direction
$l_4=\frac{1}{k_{B2}}(0,1,0)$, where
\beqs
k_{B_2}= \frac{1}{2A^2}\left|\frac{\nu + 2k}{\nu-2k}\right|^{3\over4}\left[ (R+\sigma)^2 - \sigma_2^2\right]^{-{1\over2}}\left[\frac{(R+\sigma_1)^2 -\sigma_2^2}{(R-\sigma_1)^2 -\sigma_2^2}\right]^{1\over4}.
\eeqs
It corresponds to a bolt with surface gravity $k_{B_2}$.
\item \textbf{Rod 5} - For $a_2<z<a_1$ one has a finite timelike rod, corresponding to a second black hole horizon with $S^3$ topology. Its normalized direction is
$l_5=\frac{1}{k_{H_2}}(1,0,0)$, where
\beqs
k_{H_2}=\sqrt{\frac{K_0}{2\sqrt{2}}}\frac{p_2}{A}\bigg[\left(\rho\tilde{f}^{-1}e^{\tilde{\mu}}\right)|_{H_2}\bigg]^{-\frac{3}{4}}
\eeqs
 is the surface gravity of the black hole horizon corresponding to this rod.
\item \textbf{Rod 6} - for $z>a_1$ one has a semi-infinite spacelike rod with normalized direction
\beqs
l_6=\sqrt{\frac{2\sqrt{2}}{K_0}}(0,-2AC(\sigma-\sigma_1+R),1).
\eeqs
\end{itemize}
The expressions for the surface gravity of the black holes include two constants $p_i$, $i=1,2$. They take part in the expansion of the metric function $e^{2h-2\gamma}$ near each black hole horizon
\beqs
e^{2h-2\gamma}|_{H_i}&=&(p_i)^2\sqrt{\rho}+{\cal O}(\rho),
\eeqs
and coincide with the expressions (\ref{Pi_KK}) computed for the black Saturn on KK bubbles in the previous section. The integration constant $K_0$ should also be chosen as in the previous case $K_0 = 2\sqrt{2}$  to ensure that the solution asymptotic is locally Kaluza-Klein. The constants $A$ and $C$ obey $A^2-C^2 = 1$ as already mentioned in the construction of the solution.

Certain restrictions should be imposed in addition on the solution parameters to ensure that conical and orbifold singularities do not occur. To avoid conical singularities it is necessary that the periods of the orbits of the Killing field $\partial/\partial\chi$ in the vicinity of the two bolts coincide with the length of the compact dimension at infinity $L$

\beqs\label{conical_TB}
\Delta\chi_{B_1} = \frac{2\pi}{k_{B_1}} = L, \quad~~~ \Delta\chi_{B_2} = \frac{2\pi}{k_{B_2}} = L.
\eeqs

The two equations are satisfied simultaneously if

\beqs
\frac{\sigma-\sigma_1}{\sigma + \sigma_1} = \left|\frac{\nu - 2k}{\nu+2k}\right|^{3\over2}\left[\frac{(R-\sigma_1)^2 -\sigma_2^2}{(R+\sigma_1)^2 -\sigma_2^2}\right]^{1\over2},
\eeqs

\noindent
which coincides with the condition for absence of conical singularities for the black Saturn on KK bubbles.

Another restriction should be imposed to avoid orbifold singularities at $\rho = 0$, $z=a_5$,  where two spacelike rods intersect. The solution is free of orbifold singularities if the direction vectors of the two spacelike rods $z < a_5$, $a_5 < z < a_4$ with respect to a certain basic of generators of the $U(1)\times U(1)$ isometry group are connected by a $GL(2,\mathbb{Z})$ transformation \cite{Hollands:2007}. It is convenient to choose as a basis of generators the Killing fields along which the two semi-infinite rods are directed. Then, the condition of absence of orbifold singularities results in the explicit relation

\beqs
2C(\sigma-\sigma_1+R)= \sqrt{1+C^2}\left[\frac{\left(\sigma-\sigma_1\right)\left((R + \sigma)^2 - \sigma_2^2\right)}{\sigma + \sigma_1}\right]^{1\over2},
\eeqs
which can be always satisfied by restricting the value of the constant $C$. For a regular solution the direction vectors in the rod structure acquire the form
\beqs
l_1&=&(0,2n,1),~~~l_2=(0,4n,0),~~~l_3=\frac{1}{k_{H_1}}(1,0,0),~~~l_4=(0,4n,0),\nonumber\\
l_5&=&\frac{1}{k_{H_2}}(1,0,0),~~~l_6=(0,-2n,1)\nonumber
\eeqs
where the parameter $n = AC(\sigma-\sigma_1+R)$ is related to the NUT charge $N=n/G_5$\footnote{Here $G_5$ is the gravitational constant in five dimensions.} of the solution  \cite{Nedkova:2011hx}, and according to ($\ref{conical_TB}$) it is further connected with the length of the compact dimension at infinity as $L = 8\pi n$. It is evident from this representation that in the limit when the two horizons vanish the solution reduces to the Taub-bolt instanton, trivially embedded in five dimensions. In the limit when $C=0$, the NUT charge vanishes, and we recover the black Saturn on the KK bubbles, which we described in the previous section.

\section{Conserved charges and thermodynamic properties}

The solutions, which we obtained, are characterized by three conserved charges - the electric charge $\cal Q$, the mass ${\cal M}_{ADM}$, and the gravitational tension ${\cal T}$ \cite{Traschen:2001}, \cite{Traschen:2003}. In general, they are encoded in the asymptotic behavior of the metric functions and the matter fields. The mass and the tension can be obtained by computing the generalized Komar integrals  at infinity \cite{Townsend:2001}, \cite{Yazadjiev:2009nm}

\begin{eqnarray}\label{MT}
{\cal M}_{ADM} &=&  - {L\over 16\pi G_5} \int_{S^{2}_{\infty}} \left[2i_k \star d\xi - i_\xi \star d k \right], \\
{\cal{T}} &=&  - {1\over 16\pi G_5} \int_{S^{2}_{\infty}} \left[i_k \star d\xi - 2i_\xi \star d k \right], \nonumber
\end{eqnarray}
where $\xi = \frac{\partial}{\partial t}$ is the Killing field associated with time translations, $k  = \frac{\partial}{\partial\chi}$ is the Killing field corresponding to the compact dimension, and $L$ is the length of the compact dimension at infinity.

The mass and the gravitational tension can be expressed completely in terms of the asymptotic expansions of the  metric  functions $g_{tt}$ and $g_{\chi\chi}$. The asymptotic region of the discussed solutions is located at the limit $\sqrt{\rho^2 + z^2} \rightarrow \infty$, and it is conveniently parameterized by the coordinates $\{r,  \theta \}$ connected to the canonical coordinates by the transformation
\beqs
\rho=r\sin\theta,~~~~~~z=r\cos\theta.
\label{coordinateasympt}
\eeqs
In general the metric functions possess the following behavior

\beqs
g_{tt} = -1 + \frac{c_t}{r} + \mathcal{O}\left(\frac{1}{r^2}\right), \quad~~~
g_{\chi\chi} = 1 + \frac{c_\chi}{r} + \mathcal{O}\left(\frac{1}{r^2}\right),
\eeqs
in the asymptotic region $r\rightarrow\infty$, where $c_t$ and $c_\chi$ are particular constants for each solution. Consequently, the computation of the Komar integrals ($\ref{MT}$) leads to the expressions
\beqs
{\cal M}_{ADM} &=& \frac{L}{4G_5}\left( 2c_t - c_\chi\right), \nonumber \\
{\cal T} &=& \frac{1}{4G_5}\left(c_t- 2c_\chi\right).
\eeqs

Calculating the asymptotic behavior of the double black hole solution on a KK bubble we obtain the following expressions for the mass and the gravitational tension

\beqs
{\cal M}_{ADM}&=&\frac{L}{4G_5}\big[3(M_1+M_2)+R\big],~~~~~{\cal T}=\frac{R}{2G_5},
\eeqs
where $G_5$ is the gravitational constant in five dimensions. Similarly, the mass and gravitational tension of the black Saturn on KK bubbles are given by:

\beqs
{\cal M}_{ADM}&=&\frac{L}{4G_5}\big[3(M_1+M_2)+ R +\sigma -\sigma_1\big],~~~~~{\cal T}=\frac{1}{2G_5}(R + \sigma - \sigma_1),
\eeqs
and in the case of the black Saturn on the Taub-bolt instanton we obtain

\beqs
{\cal M}_{ADM}&=&\frac{L}{4G_5}\big[3(M_1+M_2)+ (1+2C^2)(R +\sigma -\sigma_1)\big], \nonumber \\
{\cal T}&=&\frac{1}{2G_5}(1+2C^2)(R + \sigma - \sigma_1).
\eeqs

The mass and the tension  can be calculated alternatively by means of the counter-term method. A suitable counter-term is proposed in \cite{Mann:2005cx}, \cite{Kleihaus:2009ff}, which was originally obtained for the Kaluza-Klein monopole \cite{Sorkin:1983ns}, \cite{Gross:1983hb},  but is also relevant for other solutions  with equivalent asymptotics. The Kaluza-Klein monopole, or the Gross-Perry-Sorkin monopole, as it is also known, is actually a Taub-NUT instanton trivially embedded in five-dimensional spacetime by adding a flat time dimension. Consequently, it possesses the same asymptotics as the black Saturn on the Taub-Bolt instanton. The asymptotic structure of the black holes on KK bubbles solutions also belongs to the same type as it is a particular case when the fibre bundle at infinity becomes trivial.

The other conserved quantity, the electric charge, is defined by the integral

\beqs
{\cal Q} = \frac{1}{16\pi G_5}\int_{S_{\infty}}\star F_{(2)},
\eeqs
where $F_{(2)}$ is the Maxwell 2-form and the integration is performed over the boundary of the spacelike cross-section at infinity. The integral can be expressed by a sum of two identical integrals over the two black hole horizon surfaces, and can be interpreted as local charges of the two black holes ${\cal Q}_{H_1}$ and ${\cal  Q}_{H_2}$. Consequently, the electric charge is  given by

\beqs
{\cal Q}&=& {\cal Q}_{H_1} + {\cal Q}_{H_2} = \frac{\sqrt{3}L(Q_1+Q_2)}{4G_5},
\eeqs
where $Q_1$ and $Q_2$ are the local charges of the two Reissner-Nordstr\"{o}m black holes representing the seed solution.
A related characteristic is the electric potential on each black hole horizon $H_i$
\beqs
\Phi_{H_i}&=&-A_{(1)t}|_{H_i}=-\frac{\sqrt{3}}{2}\Psi|_{H_i}= \sqrt{3}\left(\frac{M_i-\sigma_i}{Q_i}\right),
\eeqs
where $\Psi$ is the electric potential for the seed solution.

A local mass of each black holes ${\cal M}_{H_i}$ can be also defined by the Komar integral ($\ref{MT}$) evaluated over the horizon surface

\begin{eqnarray}\label{MH}
{\cal M}_{H_i} = - {L\over 8\pi G_5} \int_{H_i} i_k \star d\xi = \frac{L}{2G_5}\Delta l_{H_i}.
\end{eqnarray}

\noindent
For our class of solutions it is determined by the length of the horizon rod $\Delta l_{H_i} = 2\sigma_i$. In a similar way a local mass of the KK bubbles, or the bolts in the Taub-bolt case, can be defined by integrating ($\ref{MT}$) over the fixed point sets of the Killing field $\partial/\partial\chi$

\begin{eqnarray}\label{MB}
{\cal M}_{B_i} =  {L\over 16\pi G_5} \int_{B_i} i_\xi \star d k = \frac{L}{4G_5}\Delta l_{B_i}.
\end{eqnarray}

Explicit calculation gives that it is again proportional to the length of the rod representing the bubble $\Delta l_{B_i}$, leading to\footnote {The double  black hole  solution presented in  section $3.1$ contains only a single bubble with mass given by ${\cal M}_{B_2}$.}

\begin{eqnarray}
{\cal M}_{B_1} =  \frac{L}{4G_5}(\sigma -\sigma_1), \quad~~~ {\cal M}_{B_2} =  \frac{L}{4G_5}(R-\sigma_2-\sigma_1).
\end{eqnarray}

To avoid confusion we should note that the parameters $M_1$ and $M_2$ described as the intrinsic masses of the black holes in the double Reissner-Nordstr\"{o}m seed solution are not defined purely by a Komar integral, so they do not correspond directly to ($\ref{MH}$). They include in addition an electromagnetic term such as

\begin{eqnarray}
M_i = - {1\over 8\pi} \int_{H_i} \star d\xi + \frac{1}{2}\Psi_i Q_i = \sigma_i +\frac{1}{2} \Psi_i Q_i,
\end{eqnarray}
where $Q_i$ are the local charges of the black holes and $\Psi_i$ are the restrictions of the electromagnetic potential for the seed solution on the horizons. These relations correspond to the Smarr relations in the four-dimensional seed solution \cite{Manko:2008gb}.

The local mass of each black hole ($\ref{MH}$) can be also expressed in terms of the corresponding horizon area $A_{H_i}$ and  surface gravity $k_{H_i}$, or equivalently by the black hole temperature $T_{H_i}$

\begin{eqnarray}
{\cal M}_{H_i} = \frac{1}{4\pi G_5}A_{H_i}k_{H_i} = \frac{1}{2G_5}T_{H_i}A_{H_i}.
\end{eqnarray}

The areas of the black hole horizons can be computed by using the expressions for the corresponding horizon area for the four dimensional seed solution. More precisely, for the double Reissner-Nordstr\"{o}m solution the area of each horizon can be expressed as  \cite{Manko:2008gb}:

\beqs
A^{(4)}_{H_i}&=&4\pi\sigma_i\left(\rho\tilde{f}^{-1}e^{\tilde{\mu}}\right)|_{H_i},
\eeqs

\noindent
where the quantities $\left(\rho\tilde{f}^{-1}e^{\tilde{\mu}}\right)|_{H_i}$ are given by (\ref{rhofmu}).
Consequently, the area of each black hole horizon for the five-dimensional solutions discussed in section $3$  can be written as

\beqs
A_{H_i}&=&4\pi \sigma_iL\big[(\rho\tilde{f}^{-1}e^{\tilde{\mu}})|_{H_i}\big]^{\frac{3}{4}}\left((\rho^{\frac{1}{2}}e^{2\gamma-2h})|_{H_i}\right)^{\frac{1}{2}}.
\label{area}
\eeqs

\noindent
As already mentioned, the metric function $e^{2h-2\gamma}$ possesses the following expansion near each black hole horizon
\beqs
e^{2h-2\gamma}|_{H_i}&=&(p_i)^2\sqrt{\rho}+{\cal O}(\rho),
\eeqs
where the constants $p_i$ are given by (\ref{pi}) for the double black hole on the KK bubble, and by (\ref{Pi_KK}) for the black Saturn on KK bubbles. Hence one finds the particularly simple expressions

\beqs
A_{H_1}&=&4\pi L\left(\frac{\big[(R+M_1+M_2)(M_1+\sigma_1)-Q_1(Q_1+Q_2)\big]^{3}}{(R+\sigma_1)^2-\sigma_2^2}\right)^{\frac{1}{2}},\nonumber\\
A_{H_2}&=&4\pi L\left(\frac{\big[(R+M_1+M_2)(M_2+\sigma_2)-Q_2(Q_1+Q_2)\big]^{3}}{(R+\sigma_2)^2-\sigma_1^2}\right)^{\frac{1}{2}},
\eeqs

\noindent
for the horizon areas of the double black hole solution presented in section $3.1$. For the black Saturn on KK bubbles one finds instead

\beqs
A_{H_1} &=&4\sqrt{2}\pi L \sigma_1^{1\over2}\frac{\big[(R+M_1+M_2)(M_1+\sigma_1)-Q_1(Q_1+Q_2)\big]^{3\over2}}{(R+\sigma_1)^2-\sigma_2^2}
\left[\frac{(R+\sigma)^2 - \sigma^2_2}{\sigma_1 + \sigma}\right]^{1\over2}, \nonumber \\[2mm]
A_{H_2}&=&4\pi L \frac{\big[(R+M_1+M_2)(M_2+\sigma_2)-Q_2(Q_1+Q_2)\big]^{3\over2}}{(R+\sigma_2)^2-\sigma_1^2}\left(R + \sigma_2 +\sigma\right)^{1\over2}\left(R + \sigma_2 -\sigma_1\right)^{1\over2}. \nonumber \\ \nonumber
\eeqs

The area formulae (\ref{area}) will receive modifications in the case of the black Saturn solution on the Taub-bolt instanton. The relevant expression is given by

\beqs
A_{H_i}&=&4\pi \sigma_iLA\big[(\rho\tilde{f}^{-1}e^{\tilde{\mu}})|_{H_i}\big]^{\frac{3}{4}}\left((\rho^{\frac{1}{2}}e^{2\gamma-2h})|_{H_i}\right)^{\frac{1}{2}},\nonumber\\
&=&\frac{4\pi \sigma_iLA}{p_i}\big[(\rho\tilde{f}^{-1}e^{\tilde{\mu}})|_{H_i}\big]^{\frac{3}{4}}.
\label{areaTB}
\eeqs

\noindent
where $\Sigma|_{\rho=0}=A=\sqrt{1+C^2}$ and the metric function $e^{2h-2\gamma}$ is determined again by the following expansion near the black hole horizons

\beqs
e^{2h-2\gamma}|_{H_i}&=&(p_i)^2\sqrt{\rho}+{\cal O}(\rho).
\eeqs

\noindent
Using the expressions for the constants $p_i$ ($\ref{Pi_KK}$) we obtain the horizon areas for the black Saturn on Taub-bolt instanton in the form

\beqs
A_{H_1} &=&4\sqrt{2}\pi L A \sigma_1^{1\over2}\frac{\big[(R+M_1+M_2)(M_1+\sigma_1)-Q_1(Q_1+Q_2)\big]^{3\over2}}{(R+\sigma_1)^2-\sigma_2^2}
\left[\frac{(R+\sigma)^2 - \sigma^2_2}{\sigma_1 + \sigma}\right]^{1\over2}, \nonumber \\[2mm]
A_{H_2}&=&4\pi L A \frac{\big[(R+M_1+M_2)(M_2+\sigma_2)-Q_2(Q_1+Q_2)\big]^{3\over2}}{(R+\sigma_2)^2-\sigma_1^2}\left(R + \sigma_2 +\sigma\right)^{1\over2}\left(R + \sigma_2 -\sigma_1\right)^{1\over2}. \nonumber \\ \nonumber
\eeqs

The thermodynamical characteristics of the solutions which  we discussed are not independent since they are connected by a set of geometrical identities knows as Smarr-like relations for the mass and the tension. The general form of these laws for five-dimensional charged static solutions representing configurations of black holes and KK bubbles was derived in \cite{Kunz:2008rs}

\begin{eqnarray}\label{Smarr_KK}
{\cal M}_{ADM} &=& \sum_i M_{H_i} + \sum_j M_{B_j} + \sum_i \Phi_{H_i} {\cal Q}_{H_i}, \nonumber \\
{\cal{T}}L &=& \frac{1}{2}\sum_i M_{H_i} + 2\sum_j M_{B_j},
\end{eqnarray}
where $M_{H_i}$ and $M_{B_j}$ are the local masses of the black holes and the bubbles ($\ref{MH}$)-($\ref{MB}$),  ${\cal Q}_{H_i}$ are the charges of the black holes and $\Phi_{H_i}$ are the restrictions of the electromagnetic potential on the horizons.  It is easy to prove that the obtained physical quantities for the double black holes on KK bubbles, and the black Saturn on KK bubbles satisfy ($\ref{Smarr_KK}$). The Smarr-like relations for the mass and the tension are modified in the case when the solution's asymptotic structure is only locally Kaluza-Klein ($M^4\times S^1$), but globally the spatial cross-sections at infinity represent a more general fibre bundle. In this case the relevant relations were derived in \cite{Nedkova:2011hx}

\begin{eqnarray}\label{Smarr_TB}
{\cal M}_{ADM} &=& \sum_i M_{H_i} + \sum_j M_{B_j} +  {L\over 2} N\hat\chi_H + \sum_i \Phi_{H_i} {\cal Q}_{H_i},\nonumber \\
{\cal{T}}L &=& \frac{1}{2}\sum_i M_{H_i} + 2\sum_j M_{B_j} + L N\hat\chi_H,
\end{eqnarray}
and they contain an additional term including the NUT charge of the solution $N$ and the restriction of the NUT potential on the horizons $\hat\chi_{H}$. The NUT potential $\hat\chi$ for static charged solutions of the 5D Einstein-Maxwell equations is defined by the exact 1-form \cite{Nedkova:2011hx}

\begin{equation}\label{Nut_potential}
 d{\hat\chi} = i_\xi i_k \star d k,
 \end{equation}
where $\xi$ is the asymptotically timelike Killing field,  and $k$ is the Killing field associated with the compact dimension. It is constant on the horizons and the bolts, and for continuity reasons its values on both surfaces should coincide. The NUT potential for the black Saturn on the Taub-bolt instanton can be calculated by integrating ($\ref{Nut_potential}$) on one of the semi-infinite rods $z<a_5$ or $z>a_1$. It is convenient to choose the integration constant so that the NUT-potential  vanishes at infinity, and thus we obtain

\begin{equation}
\hat\chi = \frac{AC(1-e^{4h})}{A^2 - C^2e^{4h}},
\end{equation}

\noindent
where the metric function $h$ is defined by ($\ref{hSaturnKK}$). Consequently, the  restriction of the NUT-potential on the horizons is equal to $\hat\chi_H = C/A$. Taking into account that the NUT charge of the solution is $N = \frac{AC}{G_5}(R+\sigma - \sigma_1)$, and using the expressions for the other physical characteristics which we obtained, it can be proved that ($\ref{Smarr_TB}$) are satisfied. The Smarr-like relations for the mass and the tension can be also combined into a single relation, which coincides for all the solutions we investigated

\begin{equation}
2{\cal M}-{\cal T}L  = \frac{3}{2}\sum_i M_{H_i} + 2 \sum_i \Phi_{H_i}{\cal Q}_{H_i}.
\end{equation}

\section{Conclusions}

As a general feature, in four and five dimensions, static solutions describing multi-horizon objects are plagued by unavoidable conical singularities. From a physical point of view, the presence of these conical singularities is to be expected since they are needed to balance the gravitational and electromagnetic interaction forces between the black holes, in order to keep the systems static. On the other hand, even if the solutions containing conical singularities appear to be singular at those points, such systems can still have a well-defined gravitational action \cite{Gibbons:1979nf,Costa:2000kf,Herdeiro:2010aq,Herdeiro:2009vd}. This means that such multi-black hole solutions with conical singularities might still admit a reasonable and well-defined thermodynamic description. For spaces with KK asymptotics, this has been explicitly checked in \cite{Stelea:2011fj}.

Previous work \cite{Chng:2008sr,Stelea:2011jm,Stelea:2009ur,Stelea:2011fj} showed that in the absence of rotation the electromagnetic interaction is not strong enough to balance the gravitational attraction between the black holes and the conical singularities cannot be avoided. The solutions we constructed in this paper involve black holes on KK bubbles of nothing. These KK bubbles are all static and have positive mass. The role of the KK bubble of nothing in between the black holes is to equilibrate and balance the interaction forces among black holes and keep the system static and in equilibrium. This was initially noted in \cite{Elvang:2002br,Harmark:2007md} for the vacuum case. In our work we extended the previous results in \cite{Kunz:2008rs} to show that the balance can be kept in the most general solution describing two charged black holes on a KK bubble.  We also constructed a new solution describing a static and charged black Saturn on KK bubbles and showed that this system can be equilibrated as well. Finally, we extended our considerations to black holes on the Taub-bolt instanton, which is a generalization of the Kaluza-Klein bubble with a nontrivial NUT-charge. Previous work \cite{Stelea:2012ph} showed that the double black hole on the Taub-bolt instanton can be equilibrated, even if the black holes are non-extremal. Similarly, the static black ring on the Taub-bolt instanton can be balanced even in the vacuum case \cite{Stelea:2012xg}. One should then expect that the static and charged black Saturn solution can be constructed on this background and by carefully choosing the parameters it can be rendered regular on and outside the black hole horizons. Indeed, in our work we derived this general solution and showed that in some cases it can become regular. Since the rod structure of the Taub-bolt instanton contains a finite spacelike rod, corresponding to a generalized KK bubble, one should note that the fact that these black hole systems could be equilibrated was to be expected, given the previous experience with black holes on the usual KK bubbles.
More general systems could also be considered on these KK backgrounds, such as the charged double black rings, or multi-black holes on the Kerr instanton, however, we leave the discussion of these solutions for further work.

Finally, let us note that while the charged black hole systems on KK bubbles can be mechanically equilibrated (in absence of the conical singularities the gravitational and electromagnetic forces in between the black holes can be balanced), we have not addressed so far the question of thermodynamic equilibrium, when the two black holes have the same temperature. In general, the Hawking temperatures can be the same for each black hole horizon if one takes the trivial case of equal masses and charges $M_1=M_2$ and $Q_1=Q_2$. However, the non-extremal charged black holes can have the same temperature even if the black holes are not equal. To see this more clearly, we shall provide one simple example that this is indeed possible: take for instance in the solution given in section $3.1$ the following values of the parameters $M_1=M_2=1$, $Q_2=0$ while $Q_1=\frac{\sqrt{3}(R+2)}{2R+1}$ and finally $R=10$. Then $\sigma_1=\frac{3}{7}$ while $\sigma_2=1$. It is now clear that $R>\sigma_1+\sigma_2$ (which is the condition that the two black hole horizons do not overlap) and the Hawking temperatures of the two black hole horizons in eqs. (\ref{haw1}) and (\ref{haw2}) in section $3.1$ are equal, even if the two black holes do not have the same characteristics.

\section*{Acknowledgements}

P.N. would like to thank S. Yazadjiev for discussions,  DFG Research Training Group 1620 Models of gravity for the support, and Oldenburg University for its kind hospitality. The partial financial support by the Bulgarian National Science Fund under Grant DMU-03/6 is also gratefully acknowledged. The work of C. S. was financially supported by POSDRU through the POSDRU/89/1.5/S/49944 contract.

\end{document}